\documentclass[12pt,preprint]{aastex}
\usepackage{graphicx,epsfig,wrapfig,paralist}
\usepackage{lscape}
\usepackage{rotating}
\newcommand{\bq}{\begin{equation}}
\newcommand{\eq}{\end{equation}}

\def\gtsim{\lower.5ex\hbox{$\buildrel > \over\sim$}}
\def\ltsim{\lower.5ex\hbox{$\buildrel < \over\sim$}}

\def\apjl{ApJL}
\def\apj{ApJ}
\def\apjs{ApJS}
\def\mnras{MNRAS}
\def\araa{ARAA}

\def\aap{A\&A}
\def\aaps{A\&A Suppl.}
\def\nat{Nature}

\shorttitle{H poor CSM shells}
\shortauthors{Chatzopoulos,Wheeler}
\begin{document}
\title
{HYDROGEN-POOR CIRCUMSTELLAR SHELLS FROM PULSATIONAL PAIR-INSTABILITY SUPERNOVAE WITH 
RAPIDLY ROTATING PROGENITORS}
\author{E. Chatzopoulos\altaffilmark{1} \& J. Craig Wheeler\altaffilmark{1}}
\email{manolis@astro.as.utexas.edu}
\altaffiltext{1}{Department of Astronomy, University of Texas at Austin, Austin, TX, USA.}

\begin{abstract}
 
In certain mass ranges, massive stars can undergo a violent 
pulsation triggered by the electron/positron pair instability that 
ejects matter, but does not totally disrupt the star. After one
or more of these pulsations, such stars are expected to undergo
core-collapse to trigger a supernova explosion. The mass range
susceptible to this pulsational phenomena may be as low as
50-70~$M_{\odot}$ if the progenitor is of very low metallicity and rotating
sufficiently rapidly to undergo nearly homogeneous evolution. 
The mass, dynamics, and composition of the matter ejected in the 
pulsation are important aspects to determine the subsequent 
observational characteristics of the explosion. We examine the
dynamics of a sample of stellar models and rotation rates
and discuss the implications for the first stars, for LBV-like
phenomena, and for superluminous supernovae. We find that the shells
ejected by pulsational pair-instability events with rapidly rotating
progenitors ($>$~30\% the critical value) are hydrogen-poor and helium and oxygen-rich. 
 
\end{abstract}

\keywords{Stars: evolution, stars: mass loss, stars: rotation, (stars:) supernovae: general, (stars:) 
supernovae: individual (pulsational pair-instability)}

\vskip 0.57 in

\section{INTRODUCTION}\label{intro}

Very massive stars will be subject to dynamical instability triggered
by the formation of electron positron pairs (Rakavy \& Shaviv 1967; 
Barkat, Rakavy \& Sack 1967; Fraley 1968). In some circumstances,
this instability will lead to violent contraction of the oxygen core,
ignition of the oxygen and total disruption of the star as a 
Pair-Instability Supernova (PISN). At somewhat more modest mass, the
collapse and burning will lead to the ejection of a shell of matter,
but not total disruption, a Pulsational Pair Instability Supernova (PPISN; Barkat, Rakavy \& Sack
1967; Heger \& Woosley 2002; Woosley, Blinnikov \& Heger 2007, hereafter WBH07). 
In the case of zero rotation, WBH07
determined that stars with Zero Age Main Sequence (ZAMS) 
masses in the range 95-130~$M_{\odot}$ become PPISN. 
Chatzopoulos \& Wheeler (2012; hereafter CW12)
explored the boundary between core collapse, PPISN, and PISN for the case of zero metallicity, as a 
function of the rate of rotation of the progenitor (see also 
Yoon, Dierks \& Langer 2012). CW12 checked the dynamics
of their rotating stellar evolution models by computing one-dimensional,
non-rotating hydrodynamic models to confirm that they underwent core
collapse, PPISN, or PISN. These models are not completely self-consistent
since they ignore the dynamical effects of rotation, but are reasonably
self-consistent in the sense that they map the structure of nominally
rotating but ``shellular" stellar models into spherically-symmetric
dynamic models. 

The masses of the PPISN progenitors are sensitive to
the effect of mass loss. Mass loss is a rather uncertain process in the case of very massive stars
and can happen continuously in the form of radiatively-driven winds or gravity waves (Quataert \& Shiode 2012), 
episodically via shell ejections 
and mechanically due to rapid rotation. Furthermore, mass loss is a strong function of metallicity
and higher metallicities will prevent initially massive stars from encountering pair-instability in
the core (Yoon, Langer \& Norman 2006; Langer et al. 2007). Langer et al. (2007) estimate that rapidly
rotating PISN progenitors may be possible for metallicities $Z <$~10$^{-5}$~$Z_{\odot}$ and less likely
in the local universe. Despite the small expected rate of PISN and PPISN events in the local universe,
the possibility of those events taking place in metal-poor environments is non-zero
and potential candidates have been discussed (SLSN~2007bi; Gal-Yam et al. 2009). 
In addition, we note that chemical mixing induced by rapid rotation is not the only way to make
hydrogen-poor PPISN progenitors. Hydrogen envelope stripping via stellar winds from very massive stars
is another possibility that has been discussed, although it might also require low metallicity
(Langer et al. 2007; Yoshida \& Umeda 2011). The results
presented here may be more relevant to early universe population III PPISN progenitors, but may be
used as a guideline for potentially similar local universe, low-metallicity events (Neill et al. 2011).

The dynamics can give insight into the 
expected behavior of the resulting configuration that may have direct
implications for observations of the first stars by the {\it James Webb Space
Telescope}. In addition, the dynamical ejection of shells may 
be related to the observed impulsive mass ejection associated with
luminous blue variables (Smith \& Owocki 2006; Smith et al. 2007). 
The PPISN phenomenon
may also be relevant to various manifestations of superluminous
supernovae (SLSNe; see Gal-Yam 2012 for a review). 
Some of these events display the characteristics
of Type IIn supernovae and are clearly the result of the collision 
of an underlying explosion with a dense, optically-thick
circumstellar medium (CSM; Chevalier 1982; Chevalier \& Fransson 1984; Ofek et al. 2010;
Moriya et al. 2011; Chevalier \& Irwin 2012; Chatzopoulos, Wheeler \& Vinko 2012). 
Other SLSNe show little or no hydrogen (Quimby, et al. 2011) and
little sign of circumstellar interaction. An outstanding issue
is whether a rapidly expanding hydrogen-deficient CSM would
suppress the narrow lines normally thought to accompany CSM
interaction (Blinnikov \& Sorokina 2010; Quimby 2012, private communication). Yet other
SLSNe show no hydrogen or helium, evidence for nickel and cobalt and
a light curve (LC) that could be powered by radioactive decay and
hence might be candidates for full-fledged PISN. An example is SLSN~2007bi 
(Gal Yam et al. 2009). 
While the ejecta mass for SLSN~2007bi seems adequate to 
conform to predictions for PISN, the ejecta mass of the otherwise similar SLSN~2010kd (Vinko et al. 2012, in preparation)
seems too low to satisfy this criterion. If SLSN~2010kd cannot
be a PISN, then some question arises as to whether or not
there are alternative explanations for SLSN~2007bi, for instance
the collision of a supernova with a hydrogen and helium-deficient
CSM that, as above, might be expanding sufficiently rapidly
to broaden and mute narrow emission lines. SLSN~2006oz
shows evidence for such a hydrogen-deficient CSM (Leloudas et al.
2012). 

There is thus considerable interest in understanding the
mass, dynamics, and composition of the matter that might
be ejected in PPISN events. In this paper, we present the
details of the dynamics of some PPISN events computed by
CW12. Section 2 describes our assumptions and models, 
section 3 gives the results. Finally, section 4 discusses our conclusions.

\section{MODELS}\label{mods}

To study the dynamics of PPISN events we select some of the zero metallicity models
studied by CW12 plus a 110~$M_{\odot}$ with $Z =$~10$^{-3}$~$Z_{\odot}$. 
We concentrate on the CW12 models with masses 60, 75 and 80~$M_{\odot}$
with ZAMS rotation 50\%, 50\% and 30\% the critical value, $\Omega_{crit}$, respectively,
where $\Omega_{crit} =$~$(g(1-\Gamma)/R)^{1/2}$ and $g =$~$GM/R^{2}$ is the gravitational 
acceleration at the ``surface" of the star, $G$ the gravitational constant, $M$ the
mass, $R$ the radius of the star and $\Gamma=L/L_{Ed}$ the Eddington factor where 
$L$ and $L_{Ed}$ are the total radiated luminosity and the Eddington luminosity, respectively.
All of the models were evolved from the ZAMS up to the time of maximum compression with radiatively and mechanically-driven mass loss included, right
before the core density and temperature enter the $\Gamma<$~4/3 dynamically unstable regime, with
the Modules for Experiments in Stellar Astrophysics stellar evolution code (MESA version 4298; Paxton et al. 2011).
MESA accounts for the effects of angular momentum transport and chemical mixing due to rotation
and magnetic fields as parameterized by Heger, Woosley \& Spruit (2005) based on the prescriptions
of Spruit (1999, 2002). For more details on the physics employed in the MESA models used here 
see CW12. 
Although CW12 considered both models without mass loss and models with mass loss included, 
we note that the neglect of the effects of mass loss in the evolution of some PPISN and PISN progenitor models will
lead to super-critical rotation and improper treatment of angular momentum transport that would affect our results on the composition and properties
of the ejected PPISN shells.
All zero metallicity models presented here were considered in CW12 to estimate the effect of mass loss on the
minimum ZAMS mass of rotating PISN and PPISN progenitors (dashed lines in their Figure 5). 
In addition to those models, we also considered the evolution of a 110~$M_{\odot}$ star rotating at 30\% the critical value 
with metallicity 1/1000 that of the sun in order to investigate the characteristics of PPISN in low, but non-zero, metallicity environments which
could be relevant to some SLSNe observed in metal poor galaxies.
For radiatively-driven mass loss we used the prescirptions of Glebbeek et al. (2009) and de Jager, Nieuwenhuijzen \& van der Hucht (1988).
Rotationally-induced mass loss is equal to $\dot{M}_{rot}=\dot{M}_{no-rot}/(1-\Omega/\Omega_{crit})^{0.43}$ where $\dot{M}_{no-rot}$ is
the mass loss rate in the case of zero rotation, due to the effect of radiatively driven winds (Heger, Langer \& Woosley 2000).
The characteristics of all evolved MESA progenitor models such as the final (pre-PPISN) rotation rate ($\Omega/\Omega_{crit,f}$),
radius ($R_{f}$), carbon-oxygen core gravitational binding energy ($-E_{B,f}$)
and carbon-oxygen core mass ($M_{CO,f}$) are summarized in Table 1.

The nearly hydrostatic MESA models were then mapped into the multi-dimensional, adaptive mesh refinement (AMR)
hydrodynamics code {\it FLASH} (Fryxell et al. 2000) in order to perform one-dimensional (1-D) simulations 
to follow the dynamical collapse and subsequent pulse and ejection of material as well as nucleosynthesis. 
The transition from MESA to {\it FLASH} is an operationally smooth one because the two codes employ the same 
equation of state (HELM EOS; Timmes \& Swesty 2000) and the same nuclear reaction network. In addition,
appropriate mesh refinement selections at initialization were made in {\it FLASH} in order to achieve the desired resolution 
for accurate calculation of the core compression and subsequent shock formation and core oxygen burning. The simulation
box size for all {\it FLASH} simulations was chosen to be $\sim$~10 times larger than the stellar radius of the relevant
model in order to sufficiently follow the ejected shell and determine the mass of the unbound material after the pulse
is complete. 
  
We limited our study to ZAMS rotation rates $\leq$~50\%~$\Omega_{crit}$ because the effects of higher rotation
in the hydrodynamic equilibrium of the models mapped to {\it FLASH} become especially important.
The equation for hydrostatic equilibrium for rotating stars (Lebovitz 1967; Maeder \& Meynet 2011 and
references therein) can be expressed as:
\begin{equation}
\frac{1}{\rho}\overrightarrow{\nabla}P = -\overrightarrow{\nabla}\Phi+\frac{1}{2}\Omega^{2}\overrightarrow{\nabla}(r\sin\theta)^{2},
\end{equation} 
where $\rho$ is the local density, $P$ the local pressure, $\Phi$ the gravitational potential, $\Omega$ is the local angular velocity,
$r$ is the distance from the center of the star and $\theta$ the colatitude (angular distance from the pole of the star). 
Equation 1 can be re-written as follows in the case of one dimension and across the equator ($\theta =$~$\pi/2$) and by changing
variable from $dr$ to fluid element mass $dm_{r} =$~$4\pi r^{2} dr$:
\begin{equation}
\frac{dP}{dm_{r}}= -\frac{G m_{r}}{4 \pi r^{4}}+\frac{\Omega^{2}}{4 \pi r},
\end{equation} 
where we have used $-\overrightarrow{\nabla}\Phi =$~$(G m_{r}/r^{2})\overrightarrow{r}/r$.
Now we can consider the following ratio in order to assess the effects of rotation in hydrostatic equilibrium:
\begin{equation}
\ell=\frac{\frac{\Omega^{2}}{4 \pi r}}{\mid\frac{dP}{dm_{r}}+\frac{G m_{r}}{4 \pi r^{4}}\mid}.
\end{equation}
For zero rotation ($\Omega =$~0), $\ell=$~0. A case of $\ell$ close to unity would imply that the effects of rotation are
comparable to the combined effects of gravity and internal pressure, therefore rotation should not be ignored in the hydrodynamic
calculations. In general, the larger the value of $\ell$ the more important the effects of rotation to hydrostatic equilibrium 
become. For the MESA models mapped to {\it FLASH} in the cases of ZAMS rotation of 
30\% and even 50\% the critical rotation, $\ell$ in the core was limited to
less than 0.05, with $\ell =$~0.05 the peak value for ZAMS $\Omega/\Omega_{crit} =$~0.5 and $\ell = 2 \times 10^{-4}$ representative for ZAMS
$\Omega/\Omega_{crit} =$~0.3. In more extreme cases of rotation (ZAMS $\Omega/\Omega_{crit} =$~0.8, also presented in CW12), $\ell$ becomes
close to unity and the effects of rotation cannot be ignored. Models with this very high rate of rotation collapse in a dynamical time-scale when mapped
into {\it FLASH}. The models with ZAMS rotation 30\% and 50\% the critical value that were
mapped to {\it FLASH} within the scope of this project remain stable over long time-scales 
(greater than their corresponding free-fall dynamical collapse time-scales) before a significant fraction of their cores
encounters the pair-formation regime of $\Gamma<$~4/3 and collapse leading to PPISN shell ejection.

In this project we study only the first shell ejections due to PPISNe. As WBH07 discussed, subsequent pulses may be encountered
by a massive star depending on its initial carbon/oxygen core mass. Multiple shell ejections will interact with each other
and ultimately the ejecta of the final supernova (SN) explosion will interact with them, too, resulting in several 
luminous transient events over the duration of decades up to centuries before stellar death. Since we are just performing
1-D hydrodynamic simulations, we ignore the effect of rotation on the shape of the ejected shell. 

\section{RESULTS}\label{results}

We post-processed the {\it FLASH} simulation files for the three models from CW12 discussed above as well as the 110~$M_{\odot}$, 
$Z =$~10$^{-3}$~$Z_{\odot}$ model in order to get measures 
of the mass lost due to the violent PPISN as well as to determine the physical characteristics of the shells ejected
as a function of increasing ZAMS mass and rotational velocity as well as metallicity. 
Figure 1 presents the distributions of density,
velocity and chemical composition for all models. Details
of the characteristics of the shells ejected by the first pulse in each case are given in Table 2 where
the shell mass ($M_{sh}$), shell kinetic energy ($E_{K,sh}$), typical shell velocity ($v_{sh} =$~$(2E_{K,sh}/M_{sh})^{1/2}$) 
and the total masses of 
helium ($M_{He,sh}$), carbon ($M_{C,sh}$) and oxygen ($M_{O,sh}$) within the ejected shells are presented.
The masses and kinetic energies of
the shells were calculated by determining how much mass is gravitationally unbound after the pulse was complete. We considered
the matter to be gravitationally unbound above radii for which $E_{K}+E_{int}-U_{G} >$~0, where $E_{K}$ is the kinetic,
$E_{int}$ the internal and $U_{G}$ the gravitational binding energy of the simulated material. 

We see from Table 2 that for fixed initial ZAMS rotational velocity and increasing mass, the ejected PPISN shells are more massive and have
higher kinetic energies. On the other hand, increasing rotation leads to the ejection of shells of lower mass:
$\sim$~7~$M_{\odot}$ in the case of ZAMS $\Omega/\Omega_{crit} =$~0.3 (for the 80~$M_{\odot}$ model) and 2-4~$M_{\odot}$ in the case of ZAMS $\Omega/\Omega_{crit} =$~0.5
(for the 60 and 70~$M_{\odot}$ models).
WBH07 calculated an ejected shell of 17.6~$M_{\odot}$ in the case of a non-rotating
60~$M_{\odot}$ oxygen core. This shell mass is larger than that of 
our 30\% critically rotating 80~$M_{\odot}$ model (which forms a 55~$M_{\odot}$
oxygen core mass) and much larger than that of our 50\% critically rotating 70~$M_{\odot}$ model (which forms
a 56~$M_{\odot}$ oxygen core).

Pre-SN mass loss lead to almost entirely stripped carbon-oxygen cores for the zero metallicity CW12 models with 
70~$M_{\odot}$, $\Omega/\Omega_{crit} =$~0.5 and
80~$M_{\odot}$, $\Omega/\Omega_{crit} =$~0.3. Mass loss results in differences in the overall rotationally-induced mixing efficiency, which is affected
by angular momentum loss, and differences in the final structure and composition of the progenitor star and the PPISN shell.
The typical mass of helium within the ejected PPISN shell ranges between 0.3-1.3~$M_{\odot}$ (Table 2), 
a value that is in good agreement with the results presented in Table 5 of Yoon et al. (2012).
In accordance,
the oxygen and carbon abundances in the PPISN shells are generally enhanced since the shell now probes deeper layers in the star that extend to the carbon-oxygen core.

In Figure 1 (lower panels, horizontally) we illustrate the composition of the ejected shells. In the case of moderate rotation 
(ZAMS $\Omega/\Omega_{crit} =$~0.3) the outer regions of the progenitor stars are helium rich, with
traces of oxygen and carbon present in deeper layers.
As a result, the composition of the ejected PPISN shells is predominantly He with small traces of oxygen present
in their inner parts. 
In the case of the 70~$M_{\odot}$, $\Omega/\Omega_{crit} =$~0.5 model
the PPISN shell, though still helium rich, is significantly enriched mainly
with oxygen but also with some carbon. In some cases, the oxygen mass fractions can be up to 0.5 or more. In all rotating cases, the shells
are hydrogen-poor. The outer layers of the stars after their first PPISN are even more enhanced in oxygen and carbon, therefore subsequent
shell ejections are expected to be even more oxygen-rich, potentially leading to shell collisions of oxygen-rich material.
The luminous output from this kind of CSM interaction is not necessarily going to be similar to that observed in cases of hydrogen-rich 
CSM interaction. Emission lines of hydrogen and, in some cases, of helium will be absent in the spectrum of oxygen-rich events. 

The 110~$M_{\odot}$ model, with $Z =$~10$^{-3}$~$Z_{\odot}$ and ZAMS rotation 30\% the critical value, lost the larger fraction of its initial
mass to strong radiatively driven winds combined with rotationally-induced mass loss, which left it with a completely stripped $\sim$~41~$M_{\odot}$ C/O
core, right within the range of PPISN. The PPISN pulse was followed hydrodynamically in FLASH and the relevant dynamics are detailed in Table 2
and in Figure 2 where the density, velocity and chemical composition of the unbound PPISN shell are shown at time $t \simeq$~31000~s after the pulse.
In reality, the ejected PPISN shell from this model would collide with the previously-expelled 69~$M_{\odot}$ hydrogen/helium shell from the progenitor star leading
to a potentially long-lasting SN ejecta - CSM interaction and an associated long LC duration.
The effect of progenitor metallicity in the final ejected PPISN can be seen by comparing the zero metallicity 60~$M_{\odot}$, 
$\Omega/\Omega_{crit} =$~0.5 model with the $Z =$~10$^{-3}$~$Z_{\odot}$, 110~$M_{\odot}$, $\Omega/\Omega_{crit} =$~0.3 model since both models
make C/O cores of the same mass (41~$M_{\odot}$). We find that the PPISN shell of the $Z =$~10$^{-3}$~$Z_{\odot}$ model is more significantly
enhanced in carbon and oxygen and more depleted in helium 
than the zero metallicity model mainly due to the fact that deeper layers are probed as a result
of extreme mass loss for the 10$^{-3}$~$Z_{\odot}$ model. In addition, we find a larger PPISN shell with a smaller kinetic energy
associated with the ~10$^{-3}$~$Z_{\odot}$ model. The fact that the PPISN phenomenon is possible for non-zero metallicities that
may be relevant to metal-poor dwarf galaxies means that these brilliant events may be related to some nearby, 
hydrogen-poor SLSNe such as SN~2007bi.

\section{DISCUSSION AND CONCLUSIONS}\label{disc}

 In this paper we have discussed the properties of shells ejected by massive (60-80~$M_{\odot}$), rotating (30\%-50\% the critical 
value on the ZAMS), stars with zero (and one case of 10$^{-3}$~$Z_{\odot}$) metallicity encountering PPISNe for the first time. 
We find that for increasing PPISN progenitor 
rotational velocities the resulting pulses are less energetic and shells of smaller masses but rich in helium, carbon and oxygen are ejected.
For the range of models considered here, the masses of the ejected shells vary from
$\sim$~2~$M_{\odot}$ for higher rotation values all the way up to $\sim$~7~$M_{\odot}$ for lower rotation. We find that the shells from
the first PPISN ejections are all rich in helium, oxygen and carbon in constrast to hydrogen-rich shells ejected in non-rotating cases
(WBH07). We note, however, that subsequent pulses in the case of zero rotation might also lead to helium-rich shells, since
deeper layers of the star are probed. Zero rotation models 
are not expected to lead to shells with significantly enhanced carbon and oxygen as is the
case for rotating progenitors. The ejection of hydrogen-poor shells from massive population III stars in the early universe might have
important implications for the composition of the interstellar medium in these epochs. 

 Our results imply that rotationally-induced chemical mixing (mainly due to meridional circulation and the Spruit-Tayler mechanism
for the effects of magnetic fields) in zero metallicity massive stars leads to homogeneous evolution
and larger carbon/oxygen core masses before encountering pair-instability than do non-rotating models of the same mass, 
as shown in CW12 (see also Yoon, Dierks \& Langer 2012). 
We also examined the case of a low metallicity ($Z =$~10$^{-3}$~$Z_{\odot}$), 110~$M_{\odot}$ star which produces an entirely stripped
(41~$M_{\odot}$) C/O core and encounters PPISN which leads to the ejection of a  $\sim$~3~$M_{\odot}$ shell that is significantly enhanced in carbon and oxygen. 
This model was run to indicate that the PPISN phenomenon leading to hydrogen poor ejected shells might also be relevant to low metallicity
environments such as dwarf galaxies that seem to be the host environments for some hydrogen-poor SLSNe.
The strong chemical mixing initially stirs
helium and later oxygen and carbon to the outer layers while dredging hydrogen inward to the core. 
When the
carbon/oxygen cores of those stars encounter PPISN they eject those helium and metal-enriched outer layers therefore chemically enriching
the surrounding circumstellar medium. 
Subsequent pulses may be even richer in carbon and oxygen since they probe the inner regions of the star, leading
to collisions of hydrogen-poor shells. Ultimately, the final SN explosion takes placed embedded within this chemically enriched CSM
and the SN ejecta interact with it. 

This kind of hydrogen-poor CSM interaction is not necessarily going to possess the same observational characteristics as 
hydrogen-rich CSM interaction. Hydrogen-rich CSM interaction seems to be related to Type IIn SNe, the spectra of which show narrow
emission lines of hydrogen and, sometimes, weaker emission lines of helium. 
SLSN events such as
SLSN~2006tf (Smith et al. 2008), SLSN~2006gy (Smith et al. 2007, 2010)
and SLSN~2008es (Gezari et al. 2009; Miller et al. 2009) 
seem to fall into this category. On the other hand, recent discoveries of 
SLSNe with no signs of hydrogen in their spectrum (Quimby et al. 2011; Leloudas et al. 2012)
might indicate that not all CSM interaction involves hydrogen-rich material. 
Additionally, some of those hydrogen-poor events show an early precursor plateau in their LCs 
(Blinnikov \& Sorokina 2010; Dessart et al. 2011; Leloudas et al. 2012). 

In this context, a hydrogen-poor CSM interaction might also be an alternative explanation for the nature of SLSN~2007bi (Gal-Yam et al. 2009), which
is considered the strongest observed candidate for PISN. At first, the CSM interaction model for this event was ruled out due its spectral
characteristics showing no typical signs of hydrogen-rich interaction, given the absence of narrow hydrogen lines from any of the spectra
obtained. The optical spectrum predicted for helium/carbon/oxygen-rich CSM interaction, which could result from PPISNe with rapidly rotating
progenitors, is unexplored, but it must, perforce, be free of hydrogen features. For this reason, 
Chatzopoulos et al. (2012, in preparation) will present a semi-analytical CSM interaction model fit
to the observed LC of SLSN~2007bi considering this to be a possible alternative model. 
Future multi-group radiation hydrodynamics simulations of such events are expected to shed more light on the issue.

\acknowledgments
We wish to thank the anonymous referee for valuable comments and suggestions
that helped significantly improve this paper.
We thank the MESA team for making this valuable tool readily available
and especially thank Bill Paxton for his ready advice and council 
in running the code. We thank Volker Bromm and Milos Milosavljevic 
for discussions on the topic and Sean Couch and Christopher Lindner 
for offering advice on {\it FLASH} dynamics.
This research is supported in part by NSF AST-1109801. EC wishes to thank
the University of Texas Graduate School for the William C. Powers fellowship given in support
of his studies.


{}              

\newpage                


\begin{figure}
\begin{center}
\includegraphics[angle=-90,width=18cm]{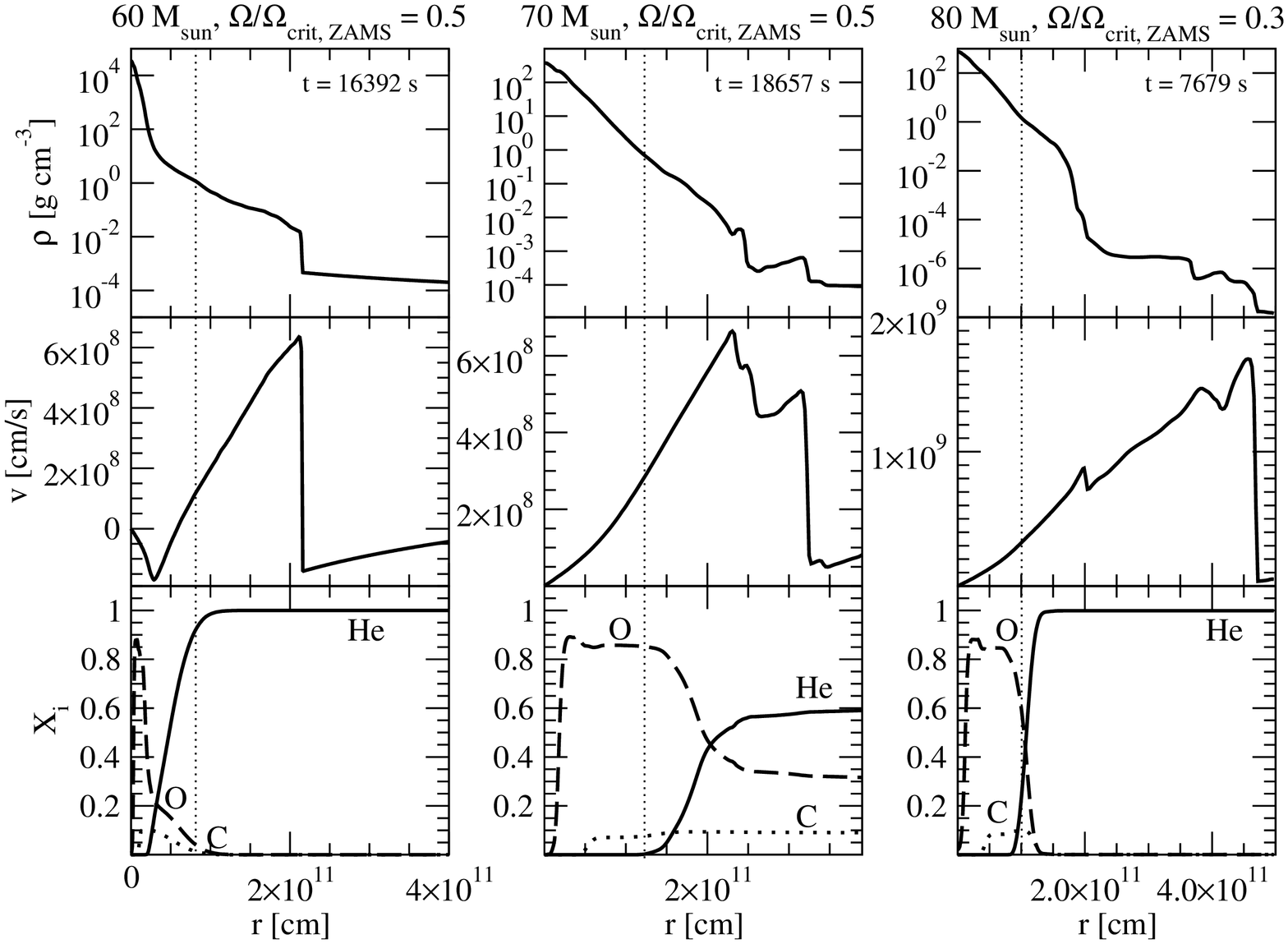}
\caption{Density (upper panels), velocity (middle panels, horizontal) and composition profiles (lower panels) for
shells ejected by PPISNe of progenitor masses 60~$M_{\odot}$ (left panels), 70~$M_{\odot}$ (middle panels, vertical) and 80~$M_{\odot}$ (right panels)
with ZAMS rotational velocities 50\%, 50\% and 30\% the critical value accordingly. 
In the composition profiles,
the solid curves show the mass fraction of helium, the dashed curves the mass fraction of oxygen and the dotted
curves the mass fraction of carbon. The time since the first PPISNe pulse in each case is given in the upper panels.
In all panels, the dashed vertical lines indicate the radii above which the 
material is gravitationally unbound.}
\end{center}
\end{figure}

\begin{figure}
\begin{center}
\includegraphics[angle=-90,width=18cm]{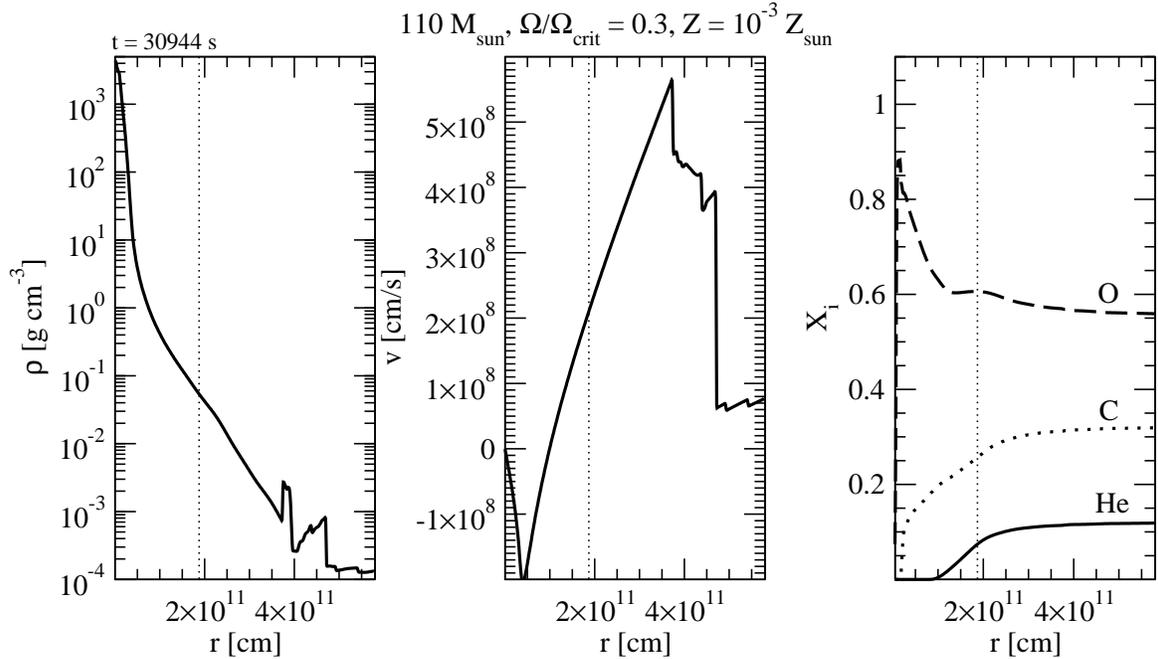}
\caption{Density (left panel), velocity (middle panel) and composition profiles (right panel) for
shells ejected by a PPISN of ZAMS progenitor masse 110~$M_{\odot}$ 
with ZAMS rotational velocitiy 30\% the critical value and metallicity $Z =$~10$^{-3}$~$Z_{\odot}$.. 
In the composition profile, the solid curve shows the mass fraction of helium, the dashed curve the mass fraction of oxygen and the dotted
curve the mass fraction of carbon. The time since the first PPISNe pulse is given in the upper panel.
In all panels, the dashed vertical lines indicate the radii above which the 
material is gravitationally unbound.}
\end{center}
\end{figure}


\clearpage
\setcounter{table}{0}
\begin{deluxetable}{llllllllccccccc}
\tabletypesize{\tiny}
\tablewidth{0pt}
\tablecaption{Physical characteristics of the pre-PPISN models used in this work.}
\tablehead{
\colhead {$M_{ZAMS}$~($M_{\odot}$)} &
\colhead {$M_{f}$~($M_{\odot}$)} &
\colhead {$\Omega/\Omega_{crit,ZAMS}$} &
\colhead {$\Omega/\Omega_{crit,f}$} &
\colhead {$R_{f}$~(10$^{11}$~cm)} &
\colhead {$-E_{B,f}$~($10^{52}$~erg)} &
\colhead {$M_{CO,f}$~($M_{\odot}$)} & \\
}
\startdata
60              & 46   &0.50 & 1.00 & 1.10 & 0.43 & 41  \\
70              & 47   &0.50 & 1.00 & 0.41 & 0.73 & 46  \\	
80              & 58   &0.30 & 1.00 & 0.49 & 1.16 & 55  \\	
110$^{\dagger}$	& 41   &0.30 & 1.00 & 0.58 & 0.40 & 41  \\
\enddata 
\tablecomments{Quantities with the subscript ``f" denote pre-PPISN values.
$\dagger$ For this model the initial metallicity was $Z =$~10$^{-3}$~$Z_{\odot}$.}
\end{deluxetable}

\setcounter{table}{1}
\begin{deluxetable}{llllllllcccccccc}
\tabletypesize{\tiny}
\tablewidth{0pt}
\tablecaption{Physical characteristics of the shells ejected by the PPISNe models discussed in this work.}
\tablehead{
\colhead {$M_{ZAMS}$~($M_{\odot}$)} &
\colhead {$\Omega/\Omega_{crit,ZAMS}$} &
\colhead {$M_{sh}$~($M_{\odot}$)} &
\colhead {$E_{K,sh}$~($10^{51}$~erg)} &
\colhead {$v_{sh}$~$^{a}$~(km~s$^{-1}$)} &
\colhead {$M_{He,sh}$~($M_{\odot}$)} &
\colhead {$M_{C,sh}$~($M_{\odot}$)} &
\colhead {$M_{O,sh}$~($M_{\odot}$)} & \\
}
\startdata
60                   & 0.5 & 1.9 & 0.25& 3636.48  & 1.71 &  0.06 & 0.13  \\
70                   & 0.5 & 3.9 & 0.31& 2826.42  & 0.23 &  0.81 & 2.86  \\
80                   & 0.3 & 7.3 & 0.48& 2570.68  & 0.32 &  0.62 & 6.36  \\   
110$^{\dagger}$      & 0.3 & 3.1 & 0.09& 1607.59  & 0.17 &  0.78 & 2.15  \\
\enddata 
\tablecomments{$^{a}$ The average shell velocity is obtained by making use of the formula $v_{sh} =$~$\sqrt{(2E_{K,sh}/M_{sh})}$.
$\dagger$ For this model the initial metallicity was $Z =$~10$^{-3}$~$Z_{\odot}$.}
\end{deluxetable}

\end{document}